\documentclass[pdflatex,sn-mathphys-num]{sn-jnl}

\usepackage{graphicx}%
\usepackage{multirow}%
\usepackage{amsmath,amssymb,amsfonts}%
\usepackage{amsthm}%
\usepackage{mathrsfs}%
\usepackage[title]{appendix}%
\usepackage{xcolor}%
\usepackage{textcomp}%
\usepackage{manyfoot}%
\usepackage{booktabs}%
\usepackage{algorithm}%
\usepackage{algorithmicx}%
\usepackage{algpseudocode}%
\usepackage{listings}%
\usepackage{natbib}

\theoremstyle{thmstyleone}%
\theoremstyle{thmstyletwo}%

\theoremstyle{thmstylethree}%

\begin{document}

\title[Article Title]{Modelling time-order effects in haptic perception with a Bayesian dynamical framework}

\author[1,2]{\fnm{Gastón} \sur{Avetta}}

\author[3]{\fnm{Jose} \sur{Lobera}}

\author[4]{\fnm{Juan José} \sur{Zárate}}

\author[1,2]{\fnm{Inés} \sur{Samengo}}

\author*[1,2]{\fnm{Damián G.} \sur{Hernández}}\email{damian.hernandez@ib.edu.ar}

\affil[1]{\orgdiv{Instituto Balseiro}, \orgname{Centro Atómico Bariloche}, \orgaddress{\country{Argentina}}}

\affil[2]{\orgdiv{Department of Applications of Physics and Biology to Health Sciences}, \orgname{Centro Atómico Bariloche}, \orgaddress{\country{Argentina}}}

\affil[3]{\orgdiv{Department of Translational Research}, \orgname{Centro Atómico Bariloche}, \orgaddress{\country{Argentina}}}

\affil[4]{\orgdiv{Institute for Intelligent Interactive Systems, Department of Computer Science}, \orgname{ETH Z\"urich}, \orgaddress{\country{Switzerland}}}


\abstract{Perceptual judgments of sequential stimuli are systematically biased by prior expectations and by the temporal structure of sensory input. In haptic discrimination tasks, these effects often manifest as time-order asymmetries, whereby the perceived difference between two stimuli depends on their presentation order. Here, we introduce a dynamical Bayesian model that accounts for these biases by combining noisy sensory measurements with an evolving internal representation of stimulus intensity. The model formalizes perception as an inference process in which prior expectations are updated by incoming stimuli and propagate in time between observations. We test the model on psychophysical data from vibrotactile discrimination experiments, in which participants compare pairs of sequential stimuli with varying intensities. With a small number of parameters, the model quantitatively reproduces both the direction and magnitude of time-order effects across subjects, as well as the observed inter-individual variability. The inferred parameters provide a compact description of perceptual biases in terms of prior expectations and noise characteristics. Beyond fitting the data, the model induces a transformation of stimulus space, leading to a subject-dependent geometry of perceived stimuli. In this transformed space, perceptual judgments exhibit approximate symmetries that are absent in the physical stimulus coordinates. These results suggest that temporal biases in perception can be understood as a consequence of dynamical inference, and that they impose non-trivial geometric constraints on perceptual representations.}




\maketitle

\section{Introduction}\label{sec1}

Sensory systems continuously integrate noisy and ambiguous information to construct a coherent percept of the world. In the tactile domain, this process is particularly dynamic: touch signals unfold in time and depend on the interaction between the body and its environment. Understanding how the brain infers tactile properties from uncertain inputs is therefore a key challenge for perceptual neuroscience and haptic technology alike.

When a tactile stimulus is delivered through a haptic device, such as a vibration, the amplitude of the signal is internally represented by neuronal activity. The transduction from external stimuli to internal representations is inherently noisy, due to uncontrolled variability in the mechanical coupling between the device and the skin, context-dependent modulatory processes in the nervous system, and electrophysiological noise in neurons and synapses. Neuronal activity therefore provides only a noisy and unreliable estimate of the true intensity of the external signal. To perform reliably under this uncertainty, the brain combines incoming sensory evidence with prior expectations about plausible stimulus values. Bayesian statistical inference offers a principled account of this process, describing how prior knowledge and noisy observations are integrated to yield a probabilistic estimate of stimulus intensity \cite{Goldreich2023,deLange2018}. This framework has been successfully applied to the modelling of perception in psychophysical experiments \citep{shi2013bayesian}, as well as to adaptation and learning in broader contexts \citep{zhou2018chance}.

In particular, discrimination thresholds have been shown to be influenced by the order in which stimuli are presented \citep{Hellstrom2003,Preuschhof2010}, a phenomenon known as \emph{time-order errors}. Previous work has highlighted the role of the inter-stimulus interval \citep{Hellstrom2004,Karim2013} and other temporal factors \citep{shi2013bayesian} in shaping this effect. Here we study a sequential haptic discrimination task \citep{Hatzfeld2017}. Our study focuses on vibrotactile stimuli, motivated by their pervasive use in mobile device notifications. Notably, early findings reported a reversal of the time-order effect at frequencies around 100–150 Hz \citep{Sinclair1996}, corresponding to the transition between Meissner and Pacinian mechanoreceptors. This frequency range is not only central to the encoding of vibrotactile signals \citep{Muschter2021} and the induction of haptic illusions \citep{Hoffmann2019}, but also underlies the operation of common actuators such as eccentric rotating mass (ERM) and linear resonant actuators (LRA).

Our central hypothesis is that subjects sustain not only an internal representation of external stimuli, but also, of the uncertainty of the representation. In other words, we assume subjects do more than just decoding the most likely stimulus: they access the entire posterior distribution of the stimulus, conditional to the noisy sensory information. When two tactile stimuli are presented sequentially, we assume subjects identify them as different whenever the internally held posterior distribution associated with the first stimulus shifts in a detectable amount due to the arrival of the second stimulus. Within this paradigm, the posterior distribution associated with the first stimulus has to be retained in memory during the inter-stimulus interval. Over this delay, the posterior is assumed to evolve, gradually drifting back toward the prior, due to a degradation of the confidence in the internal memory. As a consequence, earlier stimuli are assimilated into the prior before later ones are perceived, producing an asymmetry in how sequential inputs influence one another.

This model accounts for behavioural results in psychophysical haptic experiments, including both time-order errors and inter-subject variability. It also provides a principled way to analyse asymmetries in perceived similarity between sequential stimuli. Consistent general features can be identified despite individual differences, and falsifiable predictions can be generated for variations of the experimental paradigm. Finally, we discuss the applicability of the model to other perceptual systems and tasks.

\section{Methods}

\subsection{Behavioural paradigm}

\begin{figure}
    \centering
    \includegraphics[width=0.9 \textwidth]{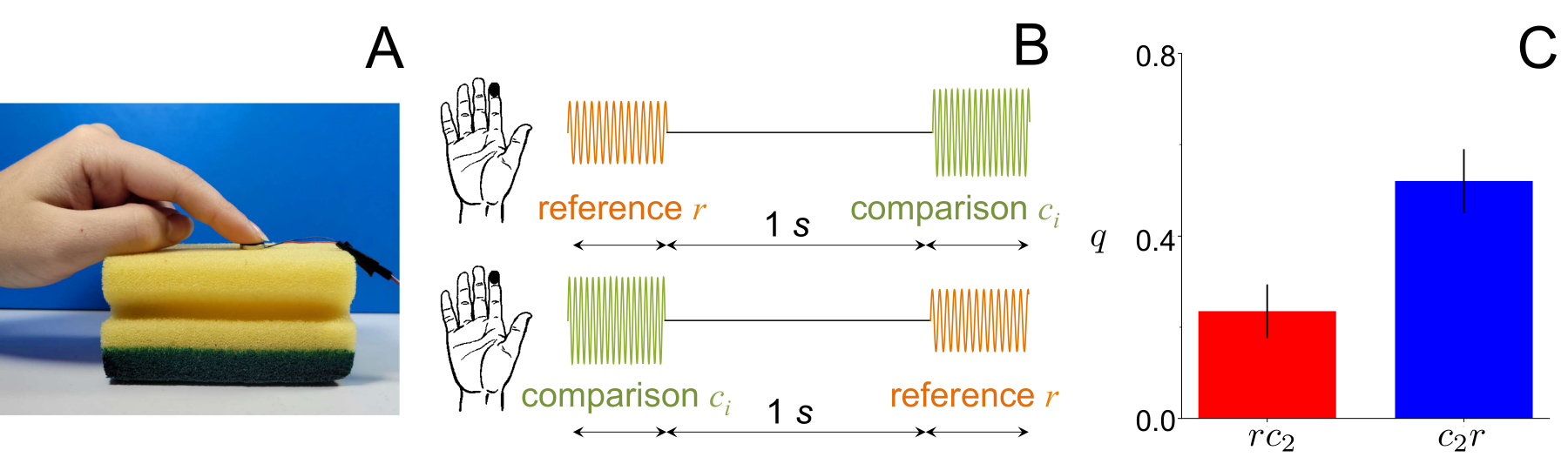}
    \caption{A: Layout of the behavioural experiments. B: The reference and each of the comparison stimuli were presented, across different trials, in the two possible orderings. C: Probabilities $q(r, c_2)$ and $q(c_2, r)$ that a given subject identify the stimuli $r$ and $c_2$ as equal, for the two presentation orderings. Red (blue): Weakest (strongest) stimulus first.  The two probabilities are significantly different ($p =  4\cdot10^{-4}$).}
    \label{fig1}
\end{figure}

To perform the vibrotactile tests, subjects positioned their index finger on a mechanical vibrating actuator rested on a sponge (Fig.~\ref{fig1}A), to minimize coupling with other rigid surfaces. Two consecutive stimuli were rendered with a 1-second pause between them, and participants informed via the keyboard whether they perceived them as equal or different (Fig.~\ref{fig1}B). For each subject and presentation order, we estimated the probability that the pair was perceived as equal. Five stimuli were used. The one with the smallest amplitude was the reference stimulus $r$, and the remaining four, the comparison stimuli $c_1, c_2, c_3, c_4$. Each test involved either a stimulus with itself, in which case the correct response was ``equal'', or the reference stimulus with one of the comparison stimuli, so the correct response was ``different''. In this case, the two possible orderings were rendered in different trials, so we were able to assess whether the probability of perceiving them as equal depended on whether the first was the strongest or the weakest. Figure \ref{fig1}C shows an example in which the ordering modulated the probability significantly.

Trials were organised in batches of $13$ tests, and the ordering of tests within a batch was randomised. Each batch contained $5$ tests of each stimulus with itself ($rr, c_1c_1, c_2c_2, c_3c_3$ and $c_4c_4$), and $2 \times 4$ tests of the reference stimulus with one of the $4$  comparison stimuli in the two possible temporal orderings ($rc_1, rc_2, rc_3, rc_4, c_1r, c_2r, c_3r, c_4r$). 

\begin{figure}
    \centering
    \includegraphics[width=0.95 \textwidth]{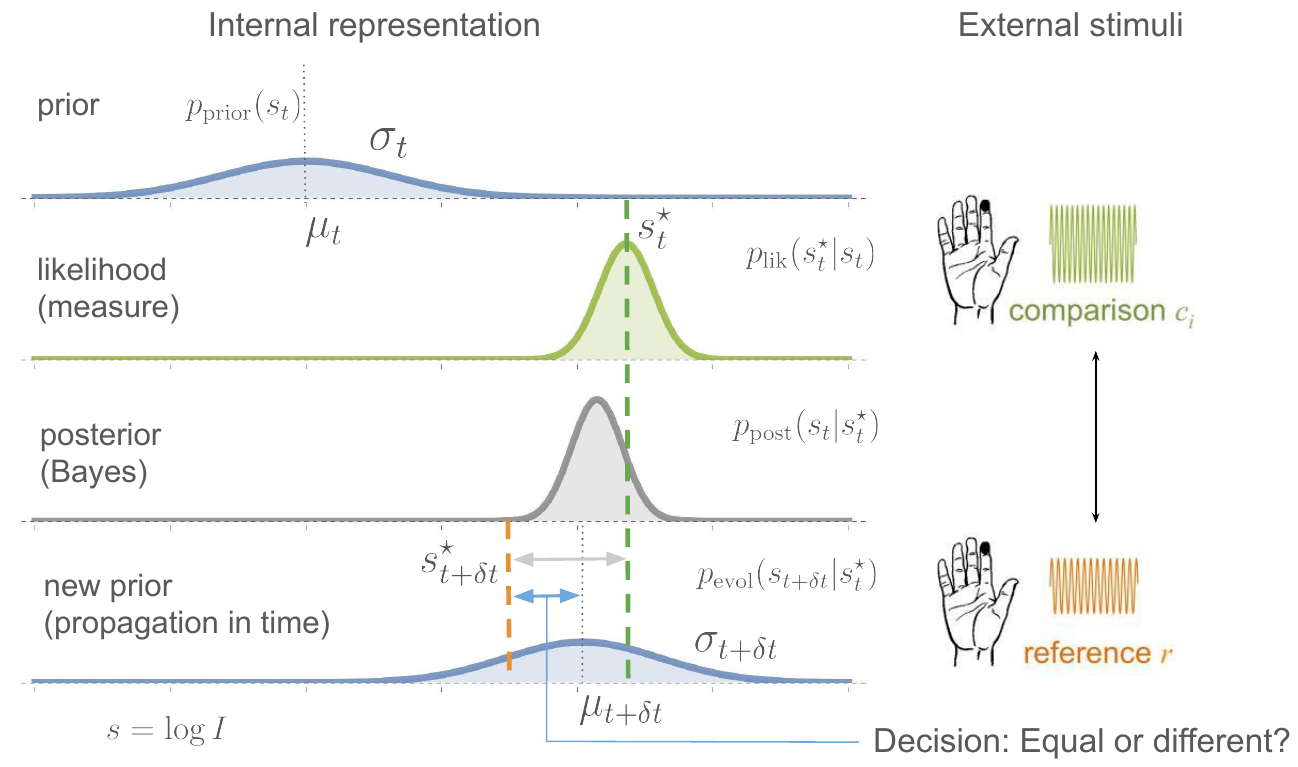}
    \caption{Bayesian model for the perception of two haptic stimuli. Before any stimulus is delivered, the subject has a prior expectation $p_\mathrm{prior}(s)$ with mean $\mu_t$ and uncertainty width $\sigma_t$. When the first stimulus of true log-intensity $s_t^\star$ is delivered at time $t$, the likelihood function, of width $\sigma_\ell$, is maximised at an estimated stimulus $s_t = s_t^\star$. The prior and likelihood are combined to produce a posterior $p_{\text{post}}(s_t|s_t^\star)$ according to Bayes' rule, centred at an intermediate point between $s_t^\star$ and $\mu_t$. In the time interval $\delta t$ between the two stimuli, the posterior gradually evolves with a propagation variance $\sigma_p^2$ towards the prior. When the second  stimulus $s_{t+\delta t}^\star$ arrives, the subject compares it with the evolved distribution, centred at $\mu_{t+\delta t}$. The probability $q$ of perceiving the stimuli as equal depends on the difference $s_{t+\delta t}^\star-\mu_{t+\delta t}$ in relation to  $\sigma_{t+\delta t}$.} 
    \label{fig2:model}
\end{figure}

To prevent discrimination to be aided by acoustic cues produced by the actuator, subjects wore noise-cancelling headphones through which white noise was played. The tests were performed by $16$ subjects, $10$ men and $6$ women, aged between $23$ and $44$ (mean $27.7$), with informed consent from the participants. All responses were treated anonymously. Subjects were na\"{\i}ve to the purpose of the study, and completed at least $50$ batches, each with $13$ comparisons.

\subsection{Design of the mechanical stimuli}

We used one of the most widely employed vibrotactile actuators: the ERM motor. A button-type vibration motor ($10 \times 2.0$ mm, Pololu) was driven by an Adafruit DRV2605L controller connected to a microcontroller board (ESP32-WROOM-32). The voltage profile of each stimulus consisted of a $70 \ ms$ linearly increasing ramp up to a plateau intensity, which depended on the stimulus. This plateau was sustained for $230\ ms$, followed by a $70\ ms$ linearly decreasing ramp back to zero. To characterize the mechanical response to the stimuli, the actuator and an accelerometer (GY-521/MPU-6050 module) were mounted on a 3D-printed support weighing $139$ grams, placed on a sponge. The maximum acceleration of the reference stimulus was $I_r = 0.087\,g$ (with $g = 9.8 \, \mathrm{m/s^2}$), and the comparison stimuli increased this value by 15\%, 36\%, 55\%, and 66\%, respectively. In ERM actuators, vibration intensity and frequency are coupled. The frequency of the reference stimulus was $108 \, Hz$, while the comparison stimuli had frequencies of $120, 131, 142$, and $151 \, Hz$.

\subsection{Modelling responses}

\label{sec:model}

In this subsection and the following, the pair of stimuli presented at times $t_1$ and $t_2$ are denoted as $(s_{t_1}^\star, s_{t_2}^\star)$, where depending on the trial, $s_i^\star$ may be the reference stimulus $r$ or a comparison stimulus $c$. The star distinguishes the stimulus actually presented from the stimulus estimated by the subject, denoted as $s$.

In line with Weber’s law~\citep{Fechner1860,KnillPouget2004}, the amplitude of vibrations is assumed to be encoded logarithmically. Before stimulation, the brain is assumed to hold a probabilistic expectation about the order of magnitude of the stimulus amplitude $I$, which we formalize as a prior distribution $p^\star(s^\star)$ over $s^\star = \log(I)$. Priors of this form have been shown to account for the \emph{central tendency} effect in other sensory modalities~\citep{shi2013bayesian}.

This work is based on three central hypotheses. The first is that, when a haptic stimulus of log-intensity $s^\star$ is delivered, the activity of tactile neurons allows the brain to produce an estimate of not only the stimulus, but also, of the uncertainty of the estimate. We model this ability by assuming that neuronal activity represents a posterior distribution $p_\mathrm{post}(s | s^\star)$.
We do not address here the detailed neural code by which the distribution is instantiated in the brain (see~\citep{Pouget2000,Ma2006,Fiser2010,Echeveste2022} for discussions of candidate mechanisms). Instead, our focus is on the behavioural implications of assuming that the brain encodes the whole posterior.

The prior and the posterior are related, through Bayes' rule, by the likelihood
\begin{align} 
    p_\mathrm{lik}(s^\star| s) &= \frac{p_\mathrm{post}(s | s^\star) p^\star(s^\star)}{p_\mathrm{prior}(s)}, \ \ \ \ \text{where} \label{eq:link} \\ p_\mathrm{prior}(s) &= \int p_\mathrm{post}(s | s^\star) \ p^\star(s^\star) \ \mathrm{d}s^\star . \label{eq:newprior}
\end{align}
The posterior distribution $p_\mathrm{post}(s|s^\star)$, when evaluated at the real $s^\star$ and seen as a function of $s$, describes the probability that the brain infers $s$ when stimulated with $s^\star$. Supplementary Appendix A derives this posterior from noisy internal representations of the stimulus. The likelihood $p_\mathrm{lik}(s^\star|s)$, evaluated at the real $s^\star$ and seen as a function of $s$, grades the inferred stimuli $s$ by their ability to make the actually delivered $s^\star$ a highly likely one. Once the prior $p^\star$ and the posterior $p_\mathrm{post}(s | s^\star)$ are known, a prior $p_\mathrm{prior}(s)$ on the estimated stimulus $s$ can be derived, as well as the likelihood $p_\mathrm{lik}(s^\star| s)$ from Eqs.~(\ref{eq:link}) and (\ref{eq:newprior}).

The second main hypothesis of our work is that posterior distributions evolve in time due to the increasing uncertainty (or forgetting) that degrades the memory of past stimuli. If the stimulus $s^\star_t$ is delivered at time $t$ and no new stimulus is rendered during the interval $\delta t$, a propagator $p_{\text{prop}}(s_{t+\delta t}|s_t)$ describes the increasing uncertainty in the representation of the stimulus due to forgetting~\citep{zhou2018chance}. The evolved posterior is hence 
\begin{align}
p_{\text{post}}(s_{t+\delta t}| s^\star) &= \int p_{\text{prop}}(s_{t+\delta t}|s_t) \ p_{\mathrm{post}}(s_t | s^\star) \ \mathrm{d}s_t \label{eq:integrales}\\ &\propto
\int  p_{\text{prop}}(s_{t+\delta t}|s_t)\, p_{\text{lik}}(s^\star|s_t)\, p_{\text{prior}}(s_t) \ \mathrm{d}s_t. \nonumber
\end{align}
The propagator formalizes the gradual reversion of the posterior toward the prior distribution, here assumed to be subject-dependent.

Many studies have used a Bayesian framework to construct the posterior $p_\mathrm{post}(s|s^\star)$, from which a so-called \emph{decoded stimulus} could be inferred. Our focus here is different. Our aim is not to decode a stimulus, but to model the processes by which the brain detects \emph{differences} between stimuli. In this framework, the third hypothesis of our work refers to the mechanism by which subjects compare sequentially delivered stimuli. When evaluating whether two stimuli rendered at times $t$ and $t + \delta t$ are different or not, we assume that subjects assess whether the evolved posterior corresponding to the first stimulus changes in a discernible amount when the second stimulus arrives. The posterior immediately before $t + \delta t$ is the updated estimation of the first stimulus, and it is here taken to be the prior for the estimation of the second stimulus. The posterior immediately after $t + \delta t$ reflects the estimation of the second stimulus that arises from a Bayesian framework that combines the new sensory information with the evolved posterior of the previous stimulus. The comparison between the two is an assessment of whether the new information displaces the posterior significantly. The evolution of the posterior of the first stimulus instantiated by the propagator during the waiting time introduces an asymmetry in the role played by the first and second stimuli, as reported below.

To simplify the mathematical description of the prior, likelihood, and propagator, we assume that  the three are Gaussian, which is closely related to a linear Kalman filter \citep{chen2003bayesian}. Non-linear effects, which are neglected here, may become significant in experiments involving a broader range of haptic stimuli \citep{zhou2018chance}. The resulting inference framework closely parallels information processing models in which an internal reference, or memory, is updated iteratively according to a linear rule \citep{shi2013bayesian,Hellstrom2003}.

Each Gaussian  is fully characterised by its mean and variance \citep{roach2017generalization,zhou2018chance}. The prior distribution $p_{\text{prior}}(s)$ has mean $\mu_t$ and variance $\sigma_t^2$. The variance $\sigma_\ell^2$ of the likelihood is associated with the uncertainty of the estimation process, since the estimated stimulus $s$ typically differs from the true stimulus $s^\star$. The precision of the estimation is limited by transduction noise due to mechanical friction, the position of the finger, and other modulatory or neuronal uncontrolled factors (see Supplementary Appendix A for the ideal-observer derivation connecting the noise in the transduction process with the posterior). The propagator is typically modelled as a diffusive process, so that its variance is expected to increase linearly with $\delta t$. We coin $\sigma_p^2$ to represent its value for the interval of the experiment $\delta t = 1$~s. The evolved distribution has mean $\mu_{t+\delta t}$ and variance $\sigma_{t+\delta t}^2$, defined by the relations (see Supplementary Appendix B)
\begin{equation}
    \begin{array}{l}
    \displaystyle \frac{\sigma_{t+\delta t}^2}{\sigma_p^2(\sigma_{t+\delta t}^2-\sigma_p^2)}=\frac{1}{\sigma_p^2}+\frac{1}{\sigma_\ell^2}+\frac{1}{\sigma_t^2}, \\\\
    \displaystyle \mu_{t+\delta t}= \frac{\sigma_{t+\delta t}^2}{\sigma_p^2}\left[\left(\frac{\sigma_\ell^{-2}}{\sigma_p^{-2}+\sigma_\ell^{-2}+\sigma_t^{-2}}\right)s^\star+\left(\frac{\sigma_t^{-2}}{\sigma_p^{-2}+\sigma_\ell^{-2}+\sigma_t^{-2}}\right)\mu_t\right].
    \end{array}
    \label{mod:update}
\end{equation}

During the interval separating the first and second stimuli within a pair, the propagator gradually distorts the posterior, which drifts towards the prior distribution. The same happens during the interval separating the second stimulus of a pair and the first stimulus of the next pair, but we assume that, for all practical purposes, this interval is sufficiently long for the posterior to have relaxed to the prior distribution before the first stimulus of the second pair arrives. For notational convenience, the time indices are therefore relabelled as $t=1$ and $t+\delta t=2$, so that $(s_t, s_{t+\delta t}) = (s_1, s_2)$.  

Within this notation, the prior distribution has mean $\mu_1$ and variance $\sigma_1^2$. The variance $\sigma_1^2$ is determined self-consistently in Eq.~(\ref{mod:update}) by imposing $\sigma_t = \sigma_{t+1} = \sigma_1$. The remaining free parameters of the model are the variances $\sigma_p^2$ and $\sigma_\ell^2$, together with the mean $\mu_1$. The updated mean $\mu_2$ and variance $\sigma_2^2$, just before the second stimulus is delivered, are then fixed by the constraints of Eq.~(\ref{mod:update}). Just after the second stimulus $s_2^\star$, the variance of the posterior $p_\mathrm{post}(s_2 | s_2^\star)$ satisfies the relation $\sigma_f^{-2}=\sigma_2^{-2}+\sigma_\ell^{-2}$, and the mean is $\mu_f=(\sigma_2^{-2} \mu_2+\sigma_\ell^{-2} s_2^\star)/\sigma_f^{-2}$.

To capture the idea that perceiving a difference amounts to detecting a change in the estimated posteriors, here we model the probability $q$ of perceiving the two stimuli of log-intensities $s_1^\star$ and $s_2^\star$ as equal (with $1 - q$ corresponding to the perception of difference) as a decreasing function of the displacement of the posteriors, the latter measured in units of the standard deviation of the evolved posterior of the first stimulus.  A large displacement $\Delta s$ (independently of the sign) should correspond to $q \to 0$. One simple model capturing this idea is
\begin{equation}
\log_2 q \propto -\left(\frac{\Delta s}{\sigma_2}\right)^2=-\left(\frac{\mu_f-\mu_2}{\sigma_2}\right)^2=-\left(\frac{s_2^\star-\mu_2}{\sigma_2+\sigma_\ell^2/\sigma_2}\right)^2
\label{mod:qdphi}
\end{equation}
or equivalently,
\begin{equation}
\displaystyle q(s_1^\star, s_2^\star) = q_0 \cdot 2^{-\left(\frac{s_2^\star-\mu_2}{\sigma_2+\sigma_\ell^2/\sigma_2}\right)^2},
\label{mod:q}
\end{equation}
where $q_0$ is a lapse parameter accounting for the fact that subjects may occasionally report two perfectly equal stimuli as different, due to distractions or fluctuations in the internal state. In this formulation, $s_1^\star$ influences the judgment only through its effect on the evolved posterior $(\mu_2, \sigma_2)$. The exponent in Eq.~\ref{mod:q} reflects that the second stimulus $s_2^\star$ is compared against the mean $\mu_2$ of the evolved distribution, rather than directly with the original stimulus $s_1^\star$, and that the distance $|s_2^\star - \mu_2|$ is normalized by the uncertainty $(\sigma_2+\sigma_\ell^2/\sigma_2)$ of the evolved posterior.

Assuming that the evolved posterior relaxes back to the original prior after each comparison, the joint log-likelihood of a subject’s responses to a series of stimulus pairs ${(s_{1j}^{\star}, s_{2j}^{\star})}$ across trials $j = 1, \dots, J$ is
\begin{equation}
\begin{split}
\mathscr{L} &=  \sum_{j \in \text{Eq}} \log \left[q\big(s_{1j}^{\star}, s_{2j}^{\star} \,\big|\, \mu_1, \sigma_p^2, \sigma_\ell^2, q_0\big) \right] \\
&\quad + \sum_{j \in \text{Diff}} \log \left[1 - q\big(s_{1j}^{\star}, s_{2j}^{\star} \,\big|\, \mu_1, \sigma_p^2, \sigma_\ell^2, q_0\big)\right],
\end{split}
\label{mod_logp}
\end{equation}
where the sets $\mathrm{Eq}$ and $\mathrm{Diff}$ contain the indices of trials reported as “equal” or “different,” respectively. In this formulation, the free parameters $(\mu_1, \sigma_p^2, \sigma_\ell^2, q_0)$ are estimated by maximizing $\mathscr{L}$.

\subsection{Estimation of model parameters and their uncertainties}

The model parameters are estimated by maximizing the log-probability. Our experiments consist of repeated presentations of a fixed set of stimulus pairs ($M = 13$ pairs in total). By grouping identical stimulus pairs in the log-probability expression, and incorporating an $L_2$ regularization term on $\mu_1$ with a broad prior variance ($V_1 = 10^2$), we obtain
\begin{equation}
\begin{split}
\mathscr{L} &= \sum_{m=1}^M  n_{\text{Eq}}^{(m)} \log \left[q\big(s_{1m}^\star, s_{2m}^{\star} \,\big|\, \mu_1, \sigma_p^2, \sigma_\ell^2, q_0\big) \right] \\
&\quad + \sum_{m=1}^M  n_{\text{Diff}}^{(m)} \log \left[1 - q\big(s_{1m}^{\star}, s_{2m}^{\star} \,\big|\, \mu_1, \sigma_p^2, \sigma_\ell^2, q_0\big) \right] \quad - \frac{\mu_1^2}{2 V_1},
\end{split}
\label{mod:logp_grouped}
\end{equation}
where $n_{\text{Eq}}^{(m)}$ and $n_{\text{Diff}}^{(m)}$ denote the number of “equal” and “different” judgements for pair $m$, respectively, with $n_{\text{Eq}}^{(m)} + n_{\text{Diff}}^{(m)} = n^{(m)}$.

\begin{figure}
    \centering
    \includegraphics[width=1.0 \textwidth]{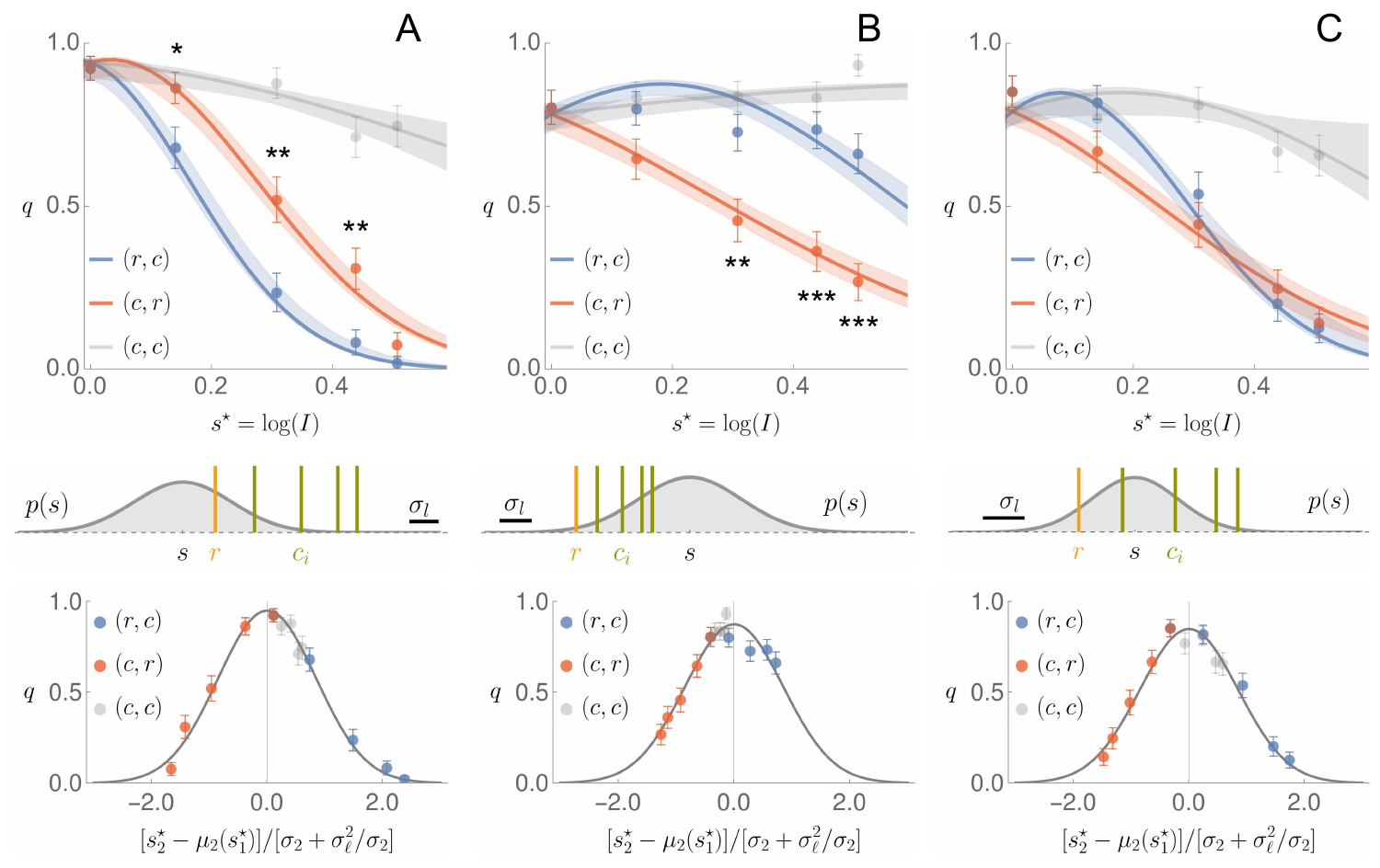}
    \caption{Top panels: Estimated probabilities $\hat{q}(r, c_i)$ (blue circles), $\hat{q}(c_i, r)$ (red circles), and $\hat{q}(c_i, c_i)$ (grey circles), of perceiving the two stimuli as equal, along with their corresponding uncertainties $\sigma_q$ (error bars, Eq.~\ref{e:2b}), plotted as a function of $s^* = \log I$, where $I$ is the amplitude of the stronger stimulus. Amplitudes are expressed in units such that $\log I_r = 0$ ($1~\mathrm{a.u.} = 0.087~\mathrm{g}$). Data are shown for three representative subjects. Solid lines with shaded areas: Maximum-likelihood fit of the Bayesian model (Eq.~\ref{mod:q}) and its confidence intervals. Middle panels: Fitted prior distribution $p_{\text{prior}}(s)$ of mean $\mu_1$, and variance $\sigma_1^2$ in relation to the reference stimulus $r$ and  comparison stimuli $c_i$. Subjects responded more accurately when the first stimulus was weaker than the second (A), less accurately (B), or with varying accuracy (C). Asterisks indicate statistically significant differences with a null hypothesis where responses collected with the two orderings of the stimuli are grouped together. Lower panels: Same probability $q$ displayed in the top panels, but now expressed as a function of $[s^\star_2 - \mu_2(s^\star_1)]/ [\sigma_2+\sigma_\ell^2/\sigma_2]$ (Eq.~\ref{mod:q}). The two orderings of the stimuli (red and blue data points, that in the upper panel fall along two different curves) now fall along one and the same continuous line.}
    \label{fig3:examples}
\end{figure}

The model parameters $\sigma_p^2$, $\sigma_\ell^2$, $\mu_1$, and $q_0$ were estimated separately for each subject by maximizing the log-likelihood $\mathscr{L}$, using the default gradient-based optimization implemented in \emph{FindMaximum} (Mathematica, Wolfram Research), under the constraint $0 < \sigma_\ell^2 < \sigma_p^2$. The fitted parameters were then substituted into Eq.~\ref{mod:q}, thereby specifying the subject-specific probability function $q(s_1^\star, s_2^\star)$. Representative examples are shown as full curves in the top panels of Fig.~\ref{fig3:examples}. Optimization converged consistently across all subjects.

Minimising the log-likelihood yields the subject-dependent parameters $\sigma_p^2$, $\sigma_\ell^2$, $\mu_1$, and $q_0$ from the observed counts $n_{\mathrm{Eq}}$ and $n_\mathrm{Diff}$. To assess the uncertainty of these optimal parameters, we note that even if the underlying function $q(s_1^\star, s_2^\star)$ remains fixed, repeated runs of the same experiment by a single subject would lead to fluctuations in $n_{\mathrm{Eq}}$ and $n_\mathrm{Diff}$ due to their binomial nature. These fluctuations, in turn, propagate as uncertainty in the estimated parameters. To quantify this effect, we sampled the distribution
\begin{equation*}
p(q \mid n_\text{Eq}) \propto p(n_\text{Eq} \mid q) \propto q^{n_\text{Eq}} (1 - q)^{n_\text{Eq}}
\end{equation*}
a large number of times ($\sim 10^3$). From each sampled $q'$, we simulated synthetic responses to obtain corresponding counts $n'_\text{Eq}$ for all stimulus pairs. Using these data, we re-estimated the model parameters $\sigma_p^2$, $\sigma_\ell^2$, $\mu_1$, and $q_0$ by minimising the log-likelihood of Eq.~\ref{mod:logp_grouped}, thereby generating a joint distribution of parameter values, represented by ellipses in Fig.~\ref{fig4:par}. For each sampled parameter set, we computed the predicted probabilities $q'(s_1^\star, s_2^\star)$ and defined confidence intervals for these curves as the central 60\% range (between the 20th and 80th percentiles), shown as shaded regions in Fig.~\ref{fig3:examples}.

In Eq.~\ref{mod:q}, the probability of judging two stimuli as equal is modelled as an exponential decay function of their relative properties, a form chosen to capture the qualitative trend observed in the experimental data. For comparison, Fig.~\ref{fig3:examples} also shows a model-free estimate $\hat{q}$ of the response probabilities (circles) together with their associated uncertainties $\hat{\sigma}_q$ (error bars), defined in terms of the sampled data as
\begin{equation} 
\hat{q} = \frac{n_\text{Eq} + 1}{n + 2},  \qquad 
\hat{\sigma}_q = \left[ \frac{n_\text{Eq} + 1}{n + 2} \left(\frac{n_\text{Eq} + 2}{n + 3}  - \frac{n_\text{Eq} + 1}{n + 2}\right) \right]^{1/2},
\label{e:2b}
\end{equation}
where $n$ is the total number of trials for the comparison in question and $n_\text{Eq}$ the number of ``equal'' responses. These expressions correspond to Bayesian estimates of the probability and its uncertainty, obtained via the Laplace estimator with a uniform prior over $[0, 1]$, which prevents degenerate results when $n_\text{Eq} = 0$ or $n_\text{Eq} = n$. For each stimulus pair, $\hat{q}$ and $\hat{\sigma}_q$ were computed separately for the two presentation orders, shown in red and blue in Fig.~\ref{fig3:examples}.

\section{Results}

\subsection{Asymmetry: Experimental results and Bayesian modelling}

In this section, the external stimuli, previously denoted by $s^\star$, are represented by the symbols $r$ and $c_i$, corresponding to the reference ($r$) and comparison ($c_i$) stimuli, respectively. 
Three representative subjects are shown in Fig.~\ref{fig3:examples}. Trials in which the weaker stimulus was presented first ($r, c$) are displayed in red, and in blue, those where the stronger stimulus went first ($c, r$). Grey curves and symbols correspond to comparisons in which both stimuli had the same intensity ($c, c$). In the upper panels, the probability of reporting the stimuli as equal is plotted as a function of the log-amplitude of the stronger stimulus $c$.

Ideal responses would be locked at $q = 1$ for equal pairs (grey) and at $q = 0$ for unequal pairs, irrespective of presentation order. Participants, however, exhibited imperfect discrimination ability, with the overall fraction of correct responses varying across the population from $59\%$ to $78\%$. As expected, the probability of judging unequal stimuli as equal converged to that of equal stimuli as the difference between them became small.

In the absence of a normative model of how comparisons are performed, no systematic asymmetry between $(r, c)$ and $(c, r)$ would be expected. Nevertheless, a subset of participants exhibited significant asymmetries, as illustrated in the upper panels of Fig.~\ref{fig3:examples}.

Some participants, such as the example in panel~A, responded more accurately when the weaker stimulus ($r$) was presented first, in which case the blue curve lies below the red one. Others showed the opposite pattern, responding more accurately when the stronger stimulus ($c$) came first (panel~B). Still others exhibited a more symmetric or mixed response (panel~C), or showed no significant difference between the two presentation orders. The Bayesian update model (solid lines) successfully captured these diverse behavioural patterns and provided a close match to the data, as the points with error bars in the upper panels of Fig.~\ref{fig3:examples} fall within the model uncertainty (shaded areas).

The response asymmetries $q(c, r) - q(r, c)$---that is, the differences between $(r, c)$ and $(c, r)$---did not vary systematically with gender (two-sample $t$-test: $p = 0.36$ for mean values, $p = 0.34$ for absolute values) or with age (Pearson correlation: $r = -0.09$). Across participants, 6 showed positive maximal asymmetries (5 significant at $p < 0.05$), and 10 showed negative maximal asymmetries (7 significant). Importantly, the magnitude of asymmetry was unrelated to overall accuracy.

The middle panels of Fig.~\ref{fig3:examples} schematically illustrate the prior distributions $p_{\text{prior}}(s)$, determined by the inferred parameters $\mu_1$ (the Gaussian mean) and the effective standard deviation $\sigma_1^2 \simeq \sigma_p^2 + \sigma_\ell^2$ (see Ec.~(\ref{eq:axis1})). For reference, the likelihood noise $\sigma_\ell$ is also indicated, together with the stimuli $r$ and $c$. When the prior distribution lies to the left of both stimuli (panel~A), the subject expects weaker estimates than those actually generated by the external stimuli. The first delivered stimulus then generates a posterior displaced to the right of the prior, with the magnitude of the shift depending on the stimulus intensity. If the first stimulus is $r$, the displacement is smaller than if it is $c$. The second stimulus is more likely to be judged ``different'' when it falls far in the tails of the evolved posterior, and this effect is particularly pronounced when the displacement is small. As a result, responses tend to be more accurate for the ordering $(r, c)$ than for $(c, r)$. Conversely, when the prior lies to the right of both stimuli (panel~B), presenting $r$ first induces a larger shift, impairing discrimination more than when $c$ is presented first. Finally, when the stimuli fall within the range of the prior (panel~C), a more symmetric or mixed response pattern emerges. Overall, the model predicts that the relative position of the prior with respect to the stimuli determines the direction of the observed asymmetry.

In our model, the probability $q$ of perceiving two stimuli as equal depends only on the normalized distance $[s_2^\star - \mu_2(s_1^\star)] /[\sigma_2+\sigma_\ell^2/\sigma_2]$, that is, on how far the second stimulus $s_2^\star$ lies from the mean $\mu_2(s_1^\star)$ of the evolved posterior of the first stimulus, expressed in units of the posterior uncertainty. To test this prediction, the lower panels of Fig.~\ref{fig3:examples} display $q$ as a function of this normalized distance. The data collapse onto a single universal curve, demonstrating that the red and blue curves in the upper panels are in fact governed by a common law.

\subsection{Uncertainty and goodness of the fitting procedure}

\begin{figure}
    \centering
    \includegraphics[width=0.85 \textwidth]{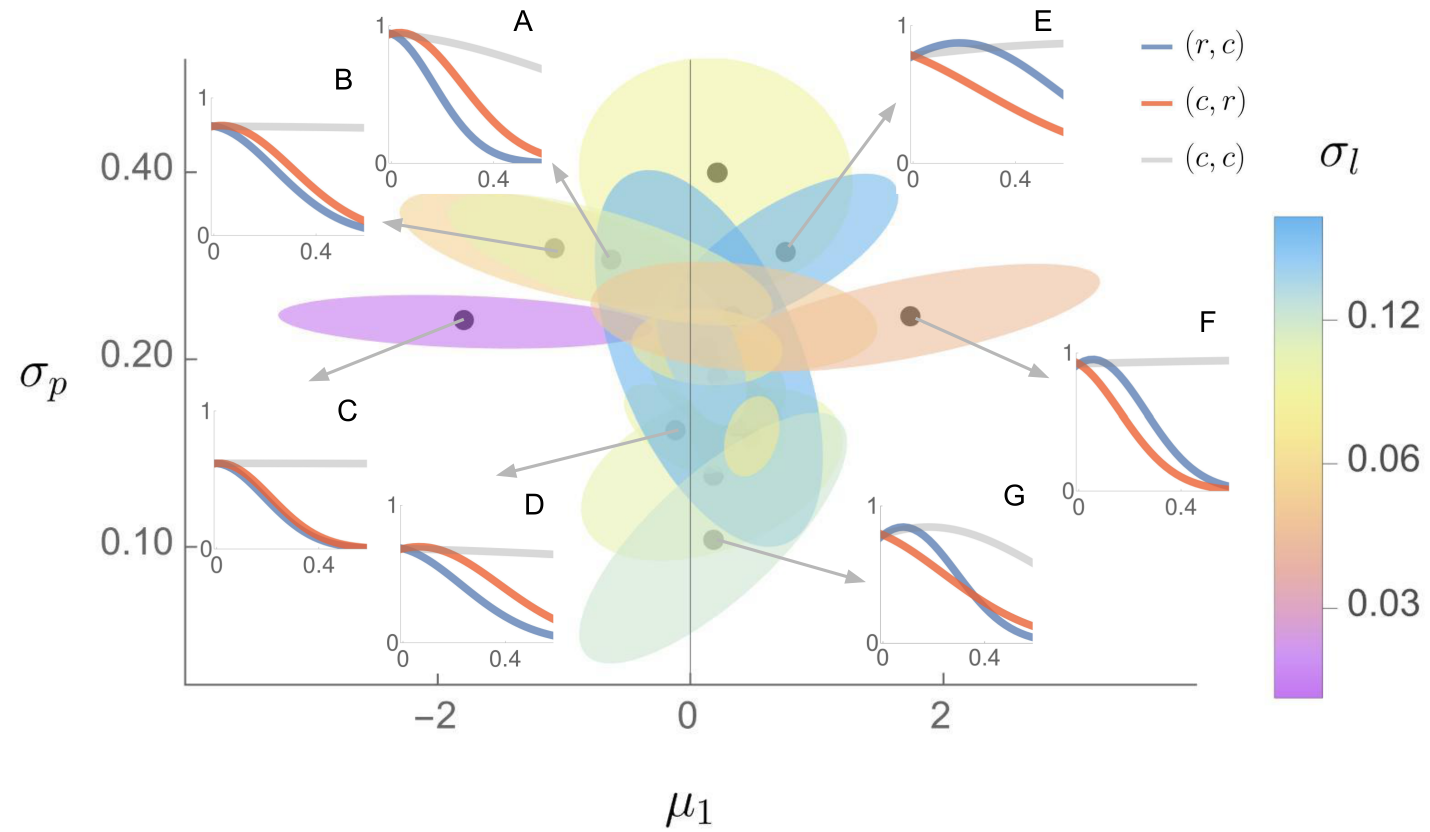}
    \caption{Inferred parameters for all the individuals in our study with prior internal representation $\mu_1$ in the horizontal axis, propagation error $\sigma_p$ on the vertical axis and likelihood noise $\sigma_\ell$ as colors. The prior uncertainty is approximately $\sigma_1^2 \sim \sigma_p^2 +\sigma_\ell^2$, mostly determined by $\sigma_p$. The ellipses represent the error in the estimation of $(\mu_1, \sigma_p)$, taking the level curve corresponding to the 20\% of the probability volume around the peak. The insets (A to G) correspond to the probability $q$ of perceiving the stimuli as equal as a function of $c=\log I$ for seven individuals (those with negative $\mu_1$ or large positive value of $\mu_1$).}
    \label{fig4:par}
\end{figure}

In Fig.~\ref{fig3:examples}, the red and blue solid lines and their shaded regions show the Maximum Likelihood fits to the measured data, providing a visually satisfactory account of the experimental results. It is important, however, to assess the goodness of fit quantitatively, to evaluate the adequacy of the proposed model, and to identify the parameters that are most important in determining $q$. In order to assess the uncertainty of the estimation, Fig.~\ref{fig4:par} displays the areas in parameter space occupying the highest 20\% of probability. The mean of the prior, $\mu_1$, emerged as the key parameter determining the direction of the asymmetry. When $\mu_1$ lay within the range of presented stimuli, responses tended to be symmetric; asymmetries were typically observed when it fell outside this range. An exception occurred when $\mu_1$ was negative but responses were nevertheless symmetric (inset C in Fig.~\ref{fig4:par}). In this case, the subject’s likelihood noise was remarkably low, $\sigma_\ell \sim 0.02$, thereby reducing the influence of the prior relative to sensory evidence. The Bayesian model was thus still able to account for the observed behaviour.

To assess the adequacy of the model, we compared the posterior evidence of the full model with that of an alternative approach in which each response probability $q$ was independently estimated using Eq.~(\ref{e:2b}) (see Supplementary Appendix C). Across all subjects, the full model—featuring a prior and an evolved posterior—was consistently favoured. Thus, the derivation presented in Sect.~\ref{sec:model}, culminating in Eq.~(\ref{mod:qdphi}), provides a more accurate and parsimonious description of the experimental data than the independent probability estimates, reducing the total number of fitted parameters within a statistically sound and conceptually unified framework. 

Quite remarkably, the predicted curves $q(s_1^\star, s_2^\star)$ remain largely unaffected by the uncertainty of the estimated parameters, as evidenced by the narrow shaded regions in Fig.~\ref{fig3:examples}. This robustness suggests that the model’s behaviour is not determined by the parameters individually, but rather by specific combinations of them. In other words, the effective dimensionality of the model could be lower than the nominal number of free parameters, pointing to the possibility of a simpler, reduced model \citep{transtrum2014model}. This reduction may be the topic of future modelling work. For the moment, however, we point out that the advantage of retaining the full formulation is that all the fitted parameters have an interpretable meaning in the framework of a Bayesian formulation. Also, a model that is not strictly minimal avoids the fine-tuning typically required by reduced models, which often suffer from the substantial complexity penalty that arises when only small regions of parameter space provide good fits \citep{mackay2003information}.

\subsection{New symmetries emerging from the model}

\begin{figure}
    \centering
    \includegraphics[width=1.\textwidth]{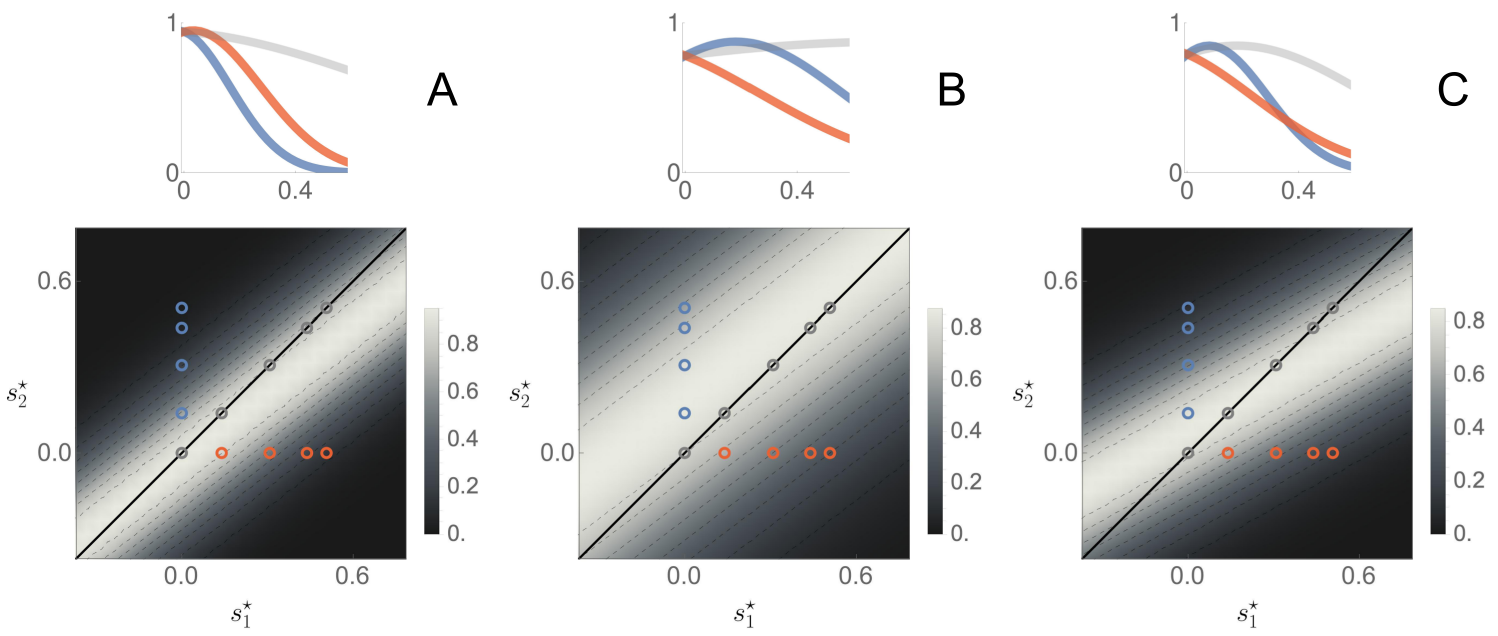}
    \caption{Symmetries of the perceptual space. Probability $q$ of perceiving two stimuli of log-intensities $(s^\star_1, s^\star_2)$ as equal. Grey level: prediction of the Bayesian perception model for three individuals  (panels A to C), same ones from Fig.~\ref{fig3:examples}. 
    Open circles: coordinates for presented stimuli in behavioural experiments.}
    \label{fig5:symm}
\end{figure}

The propagator distorts the posterior associated with the first stimulus, so the probability $q(s_1^\star, s_2^\star)$ is not symmetric with respect to an exchange of $s_1^\star$ and $s_2^\star$. However, new symmetries emerge from the model (see Fig.~\ref{fig5:symm}). These symmetries become evident when noting that it is possible to invert the order while simultaneously adjusting the intensity of one (or both) stimuli so as to leave the probability $q$ unchanged. Thus, although the bare inversion is not a symmetry of the model, an inversion accompanied by a specific change in intensity is. In Fig.~\ref{fig5:symm}, we characterize these composite symmetries and present experimental evidence supporting them.

Equation (\ref{mod:q}) implies that the probability $q(s_1^\star, s_2^\star)$ remains unchanged whenever the stimuli $s_1^\star$ and $s_2^\star$ are modified in such a way that the combination $s_2^\star - \mu_2(s_1^\star)$ does not vary. Therefore, the space of pairs of stimuli $(s_1^\star, s_2^\star)$ can be partitioned into classes of equivalence, each class corresponding to a given value for this combination. In Fig.~\ref{fig5:symm}, the classes are displayed with a constant gray level. If there were neither time-effect nor prior expectations, the gray levels should be symmetric with respect to the diagonal.

When $s_2^\star - \mu_2(s_1^\star)= 0$, the probability is maximal. Since $q(s_1^\star, s_2^\star)$ is a quadratic function of this difference, the condition $s_2^\star = \mu_2(s_1^\star)$ defines an axis of symmetry in the space $(s_1^\star, s_2^\star)$. The remaining of this section discusses the dependence of this axis on the model parameters. In Supplementary Appendix B, we show that $\mu_2(s_1^\star)$ is a linear function of $s_1^\star$, in such a way that the axis is defined by the relation 
\begin{equation}
    \displaystyle s_2^\star =\left(1-\frac{\sigma^2}{\sigma_p^2}\right)\left[\frac{\sigma^2}{\sigma_{\ell}^2} \,  s_1^\star+\,\mu_1\right],
    \label{eq:axis0}
\end{equation}
\noindent where $\sigma^2$ is the variance of the prior, and 
\begin{equation}
    \begin{array}{cc}
    \displaystyle \frac{\sigma^2}{\sigma_p^2}= G(\gamma)= \frac{1+\sqrt{1+ 4\gamma}}{2}, &\displaystyle \text{ with }\gamma=\frac{\sigma_{\ell}^2}{\sigma_p^2}.
    \end{array}
    \label{eq:axis1}
\end{equation}
Here the ratio $\gamma$ between the width of the likelihood and the variance of the propagator is the key parameter that determines both the prior uncertainty and the axis of symmetry. In fact, Eq.~(\ref{eq:axis0}) can be rewritten as
\begin{equation}
    \displaystyle s_2^\star =\left[\frac{G(\gamma)-1}{\gamma}\right] \,  s_1^\star+\,\left[\frac{G(\gamma)-1}{G(\gamma)}\right]\,\mu_1.
    \label{eq:axis2}
\end{equation}

\begin{figure}
    \centering
    \includegraphics[width=1.\textwidth]{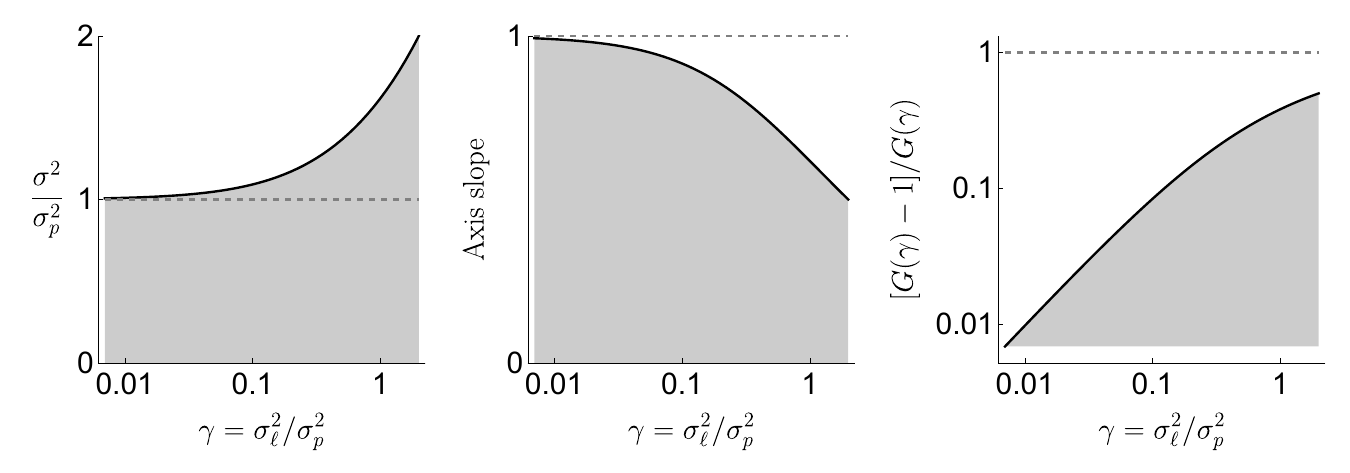}
    \caption{Properties of axis and direction of symmetry as a function of ratio  $\gamma=\sigma_{\ell}^2/\sigma_p^2$. Left panel: Prior uncertainty. Middle panel: Slope of the axis and direction of symmetry. Right panel: Prefactor of the intercept of the axis of symmetry (see Eq.~(\ref{eq:axis2})).}
    \label{fig6:gamma}
\end{figure}

Figure~\ref{fig6:gamma} shows that, when the likelihood is narrow (small $\gamma$), the slope of the axis of symmetry tends to one, and the prefactor of the intercept tends to zero. In this limit, the trivial symmetry $s_1^\star \longleftrightarrow s_2^\star$ is recovered, and time order effects are negligible. As the likelihood becomes broader, implying a looser correspondence between the real and the estimated stimuli, the perceptual and the prior uncertainty become comparable. As a consequence, the slope of the axis of symmetry diminishes, since the second stimulus is more relevant in the inference process. Moreover, the prefactor of the intercept grows, since the mean of the prior becomes more relevant.

\section{Discussion}
Through discrimination haptic experiments, we have observed that $50$\% of the tested population presented significant differences in the perception of sequential stimuli, depending on the temporal order. Approximately half of the population, for whom this asymmetry was observed, produced more accurate discriminations when the first stimulus was the strongest, and the opposite trend was observed in the other half. The tested stimuli mainly involve afferent pacinian AF-II. The population variability suggests that more than one mechanism may be in play, which may stem from response properties of the receptors (for example, a saturation effect induced by the strongest stimulus, or adaptation processes within the sensory neurons), or from cognitive strategies in the conscious process with which comparisons are made.

We built a model that coarse-grains and effectively incorporates these mechanisms in the form of a posterior distribution of the inferred intensity of haptic stimuli. The inferred distribution is sensitive to observations and also, to the passage of time, following a Bayesian inference framework with a time propagator. Within this framework, the second haptic signal is not compared to the first one, but to a shifted posterior distribution. Our model with only four parameters per subject is able to accurately fit the experimental data, explaining the variability in the population. Asymmetric responses correspond to prior expectations to the left or to the right of the range of presented stimuli, combined with an uncertain measuring process, described by the width of the likelihood function.

The naive expectation that two sequential stimuli are perceived similarly, independently of the temporal order, is broken when these internal biases and mechanisms are taken into account. In fact, our model predicts that the space of pairs of sequential haptic stimuli has specific symmetries, which are subject-dependent. In particular, a rigid transformation of the stimulus space $(s_1^\star, s_2^\star)$ into $(\tilde{s}_1^\star, \tilde{s}_2^\star)$ induces a symmetry in the probability of discrimination under $\tilde{s}_1^\star \longleftrightarrow \tilde{s}_2^\star$. This rigid transformation corresponds to taking the reflection axis given by Eq.~(\ref{eq:axis2}) to the diagonal. Moreover, the probability of distinguishing sequential stimuli remains constant under $(\tilde{s}_1^\star + \alpha, \tilde{s}_2^\star+\alpha)$, being $\alpha$ a constant magnitude. This transformation $(\tilde{s}_1^\star, \tilde{s}_2^\star)$ depends naturally on the two main ingredients of our model, the internal prior expectation and the propagation in time. If the prior is broad and close to the range of presented stimuli, and if the propagator variance is small, then time order effects are less likely to occur, corresponding to $(s_1^\star, s_2^\star) \sim (\tilde{s}_1^\star, \tilde{s}_2^\star)$. If the propagator variance increases, for example, due to an increase in the time interval between comparisons, then the transformation becomes only dependent on the last observation. The first stimulus is hence forgotten and the second is compared with the prior expectation.

We speculate that this type of model can be useful when describing more complex sequences of stimuli, which include more than two observations with varying time intervals. This type of scenario is quite likely to happen in haptic devices where signals are presented in a continuous stream to users \citep{Hatzfeld2017}. Phenomena such as adaptation and learning will modify how subjects interpret haptic signals, and a model with components such as inference and propagation in time will be necessary to understand how these signals are perceived.

\section{Conclusions}

Our results demonstrate that the so-called time-order effect in haptic perception can be quantitatively explained by a simple Bayesian inference process with dynamically evolving priors. Within this framework, biases in sequential comparisons arise not from fixed perceptual asymmetries but from the temporal evolution of expectations between successive stimuli. This perspective provides a principled computational account for perceptual asymmetries that have often been treated as empirical curiosities.

More broadly, our findings support the view that perception is an active inferential process in which the brain continuously updates predictions about the external world under uncertainty. The same mechanisms that produce the time-order effect in touch may underlie other sequential or memory-dependent perceptual biases — such as adaptation aftereffects, serial dependence, and contextual illusions — across sensory modalities.

From a cognitive standpoint, the model implies that perceptual geometry itself is not static but depends on the dynamics of expectation. As priors evolve, the metric of similarity between stimuli is reshaped, leading to systematic distortions in perceived equality. This dynamic geometry provides a unifying description of how temporal structure and probabilistic inference jointly determine perception.

Finally, the approach introduced here offers a methodological contribution: a generative Bayesian model that can be fitted to individual data and used to disentangle noise, bias, and prior evolution in perceptual judgments. Such models can help bridge behavioural data, neural mechanisms, and theoretical principles of predictive cognition.

\backmatter








\section*{Declarations}


\begin{itemize}
\item Funding. Not applicable.
\item Conflict of interest. Not applicable.
\item Ethics approval and consent to participate: All subjects participating in the behavioral experiments gave their written informed consent, as approved by Instituto Balseiro's ethics committee. The study was performed in accordance with the ethical standards as laid down in the 1964 Declaration of Helsinki and its later amendments. 
\item Materials availability. Not applicable.
\item Code availability. Not applicable.
\item Author contribution. G.A. collected and analysed data, prepared figures, J.L. provided technical and experimental support, J.J.Z. formulated the questions and contributed with the haptics design, I.S. formulated the questions, contributed with the inference approach, and participated in writing, D.G.H. formulated the questions, analysed the data, prepared figures, contributed with the inference approach, and wrote the final version of the manuscript.
\end{itemize}

\noindent

\begin{appendices}

\section{Posterior distribution and transduction noise}\label{app0}

When a stimulus of true (and unknown) log-intensity $s^\star$ is delivered, it generates noisy neural activity $\phi$. Here, $\phi$ represents the state of a collection of neurons that respond to haptic stimulation. The likelihood $p(\phi|s^\star)$ captures the variability introduced by uncontrolled factors that make the internal variable $\phi$ a noisy representation of the true stimulus $s^\star$, as for example, variability in the mechanical coupling between the device and the skin, contextual fluctuations affecting modulatory processes in the nervous system, and electrophysiological noise in neurons and synapses. As before, the brain is assumed to represent a whole distribution $p(\phi|s^\star)$. Moreover, each state $\phi$ can be used by the nervous system to infer a posterior distribution $p(s | \phi)$, where $s$ is measured in the same units as the stimulus. Marginalising over  $\phi$, the distributions $p_{\mathrm{post}}(s|s^\star)$ and $p_\mathrm{lik}(s^\star | s)$ used throughout the paper are defined by the equations
\begin{align}
    p_{\mathrm{post}}(s|s^\star) &= \int p(s | \phi) \ p(\phi|s^\star) \ \mathrm{d}\phi \nonumber \\ 
    &= \int \frac{p(\phi| s) \ p(s)}{p(\phi)} \ p(\phi|s^\star) \ \mathrm{d}\phi \label{eq:actualcalculation} \\
    &\equiv \frac{p(s) \ p_\mathrm{lik}(s^\star | s)}{p^\star(s^\star)}, \nonumber
\end{align}
where we have defined
\begin{align*}
    p(\phi) &= \int p(\phi| s) \ p(s) \ \mathrm{d}s  \\
    p_\mathrm{lik}(s^\star | s) &= p^\star(s^\star) \ \int \frac{p(\phi| s) \ p(\phi|s^\star) } {p(\phi)} \ \ \mathrm{d}\phi  
\end{align*}
This implies that if both the prior distribution $p^\star(s^\star)$, describing the statistics of external stimuli, and the noise model $p(\phi \mid s^\star)$, characterizing the variability of neural responses to a given stimulus, are known, then the posterior distribution $p_\mathrm{post}(s \mid s^\star)$ can be computed. In this paper, we do not address the neural mechanisms by which the brain might implement such a computation. We rather adopt a pragmatic stance, assuming that the calculation can, in principle, be performed, and that Eq.~(\ref{eq:actualcalculation}) is the formula that produces the posterior distribution from the point of view of an ideal-observer framework (e.g., \citep{KnillRichards1996}).

\section{Mean and variance of the updated prior}\label{app1}

Starting from the equation for the evolved posterior, renaming the indices $t\equiv 1$ and $t+\delta t\equiv 2$,

\begin{equation}
p_2(s_2) \propto \int ds_1 \;
p_{\mathrm{prop}}(s_2 | s_1) \;
p_{\mathrm{lik}}(s_1^\star | s_1) \;
p_1(s_1),
    \label{eq:app:evol}
\end{equation}
\noindent where
\begin{equation}
\begin{array}{rl}
     \log p_1(s_1) \propto &\displaystyle  -\frac{1}{2} \left( \frac{s_1 - \mu_1}{\sigma_1} \right)^2, \\\\
     \log p_{\mathrm{lik}}(s_1^\star | s_1) \propto &\displaystyle -\frac{1}{2} \left( \frac{s_1^\star - s_1}{\sigma_\ell} \right)^2, \\\\
     \log p_{\mathrm{prop}}(s_2 | s_1) \propto &\displaystyle -\frac{1}{2} \left( \frac{s_2 - s_1}{\sigma_p} \right)^2.\\
\end{array}
    \label{eq:app:logps}
\end{equation}

Inside the integral, $\int ds_1 \exp[-F(s_1,s_2)/2]\times \dots$, let us extract in $F(s_1,s_2)$ all the terms that depend on $s_1$ and $s_2$,

\begin{equation}
\begin{array}{rl}
     F(s_1,s_2) = &\displaystyle \frac{(s_1 - \mu_1)^2}{\sigma_1^2}
+ \frac{(s_1^\star - s_1)^2}{\sigma_\ell^2}
+ \frac{(s_2 - s_1)^2}{\sigma_p^2}\\\\
    = &\displaystyle s_1^2
\left(
\frac{1}{\sigma_1^2} + \frac{1}{\sigma_\ell^2} + \frac{1}{\sigma_p^2}
\right)
- 2 s_1
\left(
\frac{\mu_1}{\sigma_1^2} + \frac{s_1^\star}{\sigma_\ell^2} + \frac{s_2}{\sigma_p^2}
\right)
+ \frac{s_2^2}{\sigma_p^2} + \dots
\end{array}
    \label{eq:app:Fff}
\end{equation}

Defining
\begin{equation}
\begin{array}{cc}
    \displaystyle \frac{1}{\sigma_x^2} \equiv
\frac{1}{\sigma_1^2} + \frac{1}{\sigma_\ell^2} + \frac{1}{\sigma_p^2}, & 
\displaystyle\quad b(s_2) \equiv
\frac{\mu_1}{\sigma_1^2} + \frac{s_1^\star}{\sigma_\ell^2} + \frac{s_2}{\sigma_p^2},
\\
\end{array}
    \label{eq:app:ssbb}
\end{equation}

\noindent then we can rewrite Eq.~(\ref{eq:app:Fff}) as,
\begin{equation}
\begin{array}{rl}
F(s_1,s_2) =&\displaystyle \frac{1}{\sigma_x^2}
\left[
s_1^2 - 2 s_1 \sigma_x^2 b(s_2)
+ \left(\sigma_x^2 b(s_2)\right)^2
- \left(\sigma_x^2 b(s_2)\right)^2
\right]
+ \frac{s_2^2}{\sigma_p^2} + \dots
\\\\
=&\displaystyle \frac{1}{\sigma_x^2}
\left[ s_1 - \sigma_x^2 b(s_2) \right]^2
- \sigma_x^2 b(s_2)^2+ \frac{s_2^2}{\sigma_p^2} + \dots
\end{array}
    \label{eq:app:Fff2}
\end{equation}

After the integration over $s_1$, only the second and third terms will remain in the prior $p_2(s_2)$, all the rest being constants in $s_2$,
\begin{equation}
\begin{array}{rl}
\log p_2(s_2) \propto &\displaystyle
- \frac{s_2^2}{2\sigma_p^2}
+ \frac{\sigma_x^2}{2} \, b(s_2)^2
\\\\
= &\displaystyle -\frac{s_2^2}{2\sigma_p^2}
+ \frac{\sigma_x^2}{2} \left(
\frac{\mu_1}{\sigma_1^2}
+ \frac{s_1^\star}{\sigma_\ell^2}
+ \frac{s_2}{\sigma_p^2}
\right)^2
\end{array}
    \label{eq:app:logp2}
\end{equation}

As the new prior is also a Gaussian, we only need its mean $\mu_2$ and variance $\sigma_2^2$, which are given by
\begin{equation}
\begin{array}{c}
\displaystyle 
\frac{\sigma_2^2}{\sigma_p^2} = \frac{\sigma_p^2}{\sigma_p^2 - \sigma_x^2},
\\\\
\displaystyle \mu_2 = \frac{\sigma_2^2}{\sigma_p^2}
\left( \frac{\sigma_1^{-2}}{\sigma_x^{-2}}\mu_1 + \frac{\sigma_\ell^{-2}}{\sigma_x^{-2}}s_1^\star \right)
= \left( \frac{\sigma_p^2}{\sigma_p^2 - \sigma_x^2} \right)
\left[ \frac{\mu_1}{\sigma_1^2} + \frac{s_1^\star}{\sigma_\ell^2} \right],
\end{array}
    \label{eq:app:m2v2}
\end{equation}

\noindent where $\sigma_x^{-2}=\sigma_p^{-2}+\sigma_\ell^{-2}+\sigma_1^{-2}$.

\section{Model comparison: Predictive update vs fitting frequencies}\label{app2}

Here, we quantify whether our model with the predictive update rules is better at explaining the observed data than merely fitting the frequencies $q_{\epsilon=1}$ for each comparison independently. 

It is important to notice that we can not just compare the probability (or log-likelihood) of the data given the parameters, as our model only has four ($4$) parameters, while fitting the frequencies has thirteen ($13$) parameters. To properly compare the models, we can approximate the \emph{posterior evidence} of both hypotheses, which not only considers the optimization of the log-probability but also naturally penalizes complexity \citep{mackay2003information}. Given data $n$, parameters $\theta$ and hypothesis $\mathcal{H}_i$, the posterior evidence corresponds to $p\left(\mathcal{H} _i \mid n\right)\propto p\left(n\mid \mathcal{H}_i\right) p\left(\mathcal{H}_i\right)$, where the probability of the data given the hypothesis involves an integral over the space of parameters,
\begin{equation}
\displaystyle p\left(n\mid \mathcal{H}_i\right)  = \int d\theta\,p(\theta, n  \mid\mathcal{H}_i)  =\int d\theta\, p(n \mid \theta, \mathcal{H}_i)\,p(\theta \mid \mathcal{H}_i),
    \label{eq:app:evidence}
\end{equation}
\noindent where $p(n \mid \theta, \mathcal{H}_i)$ is the log-likelihood and $p(\theta \mid \mathcal{H}_i)$ is the prior of the corresponding hypothesis. Considering the posterior can be locally approximated around its maximum $\theta_0$ as a Gaussian with covariance matrix $\Sigma$, Eq.~(\ref{eq:app:evidence}) corresponds to
\begin{equation}
    \displaystyle \log p\left(n\mid \mathcal{H}_i\right) \simeq \log\left[p(n \mid \theta_0, \mathcal{H}_i)\,p(\theta_0 \mid \mathcal{H}_i)\right]+\log|\Sigma|^{1/2}+\frac{k}{2}\log(2\pi).
    \label{eq:app:evidapprox}
\end{equation}

\begin{wrapfigure}{r}{0.5\textwidth}
  \vspace{-20pt}
  \begin{center}
    \includegraphics[width=0.48\textwidth]{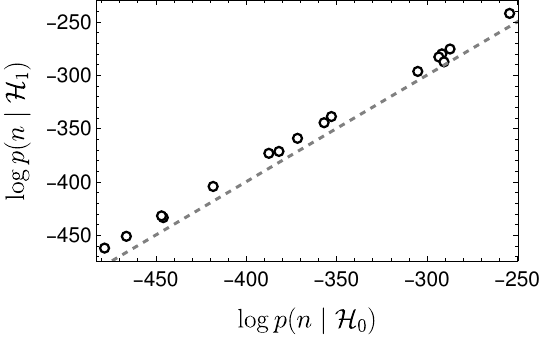}
  \end{center}
  \vspace{-10pt}
  \caption{Posterior evidence of predictive update model ($\mathcal{H}_1$) vs fitting probabilities of perceiving as equal separately ($\mathcal{H}_0$), for all $16$ subjects.}
\end{wrapfigure}

The square root of the determinant of the covariance matrix corresponds to the volume in parameter space that fits the data. For our predictive update model, we obtain the covariance matrix from the synthetic data, while for the frequencies we use the Laplace estimator to approximate this volume, $|\Sigma|^{1/2}\simeq \prod_i \hat{\sigma}_{q_i}$.

For all $16$ experimental subjects, the posterior evidence is larger for the predictive update model. Such a result is expected, as the data is fitted equally well in both cases, but the update model uses fewer parameters ($4$ vs $13$).




\end{appendices}

\bibliography{sn-bibliography}

\end{document}